\documentclass[12pt]{article}

\usepackage{times}
\usepackage{epsfig,amsmath}
\usepackage{graphicx}
\usepackage{color}
\usepackage{epstopdf}

\newenvironment{sciabstract}{%
\begin{quote} \bf}
{\end{quote}}

\newcounter{lastnote}

\title{Multipartite Entanglement of Billions of Motional Atoms Heralded by Single Photon}

\author
{Hang Li,$^{1,2,\dagger}$ Jian-Peng Dou,$^{1,2,\dagger}$ Xiao-Ling Pang,$^{1,2}$ Chao-Ni Zhang,$^{1,2}$ \\
Zeng-Quan Yan,$^{1,2}$ Tian-Huai Yang,$^{1,2}$ Jun Gao,$^{1,2}$ Jia-Ming Li,$^{3,\ast}$ Xian-Min Jin$^{1,2,\ast}$\\
\\
\normalsize{$^1$Center for Integrated Quantum Information Technologies (IQIT), School of Physics}\\ \normalsize{and Astronomy and State Key Laboratory of Advanced Optical Communication Systems}\\\normalsize{and Networks, Shanghai Jiao Tong University, Shanghai 200240, China}\\
\normalsize{$^2$CAS Center for Excellence and Synergetic Innovation Center in Quantum Information and}\\
\normalsize{Quantum Physics, University of Science and Technology of China, Hefei, Anhui 230026, China}\\
\normalsize{$^3$School of Physics and Astronomy, Shanghai Jiao Tong University, Shanghai 200240, China}\\
\normalsize{$^\dagger$These authors contributed equally to this work}\\
\normalsize{$^\ast$E-mail: lijm@sjtu.edu.cn}\\
\normalsize{$^\ast$E-mail: xianmin.jin@sjtu.edu.cn}\\
}

\date{}

\begin{document}
\baselineskip24pt

\maketitle

\begin{sciabstract}
Quantum entanglement is of central importance to quantum computing, quantum metrology, quantum information as well as the nature of quantum physics. Quantum theory does not prevent entanglement from being created and observed in macroscopic physical systems, in reality however, the accessible scale of entanglement is still very limited due to decoherence effects. Recently, entanglement has been observed among atoms from thousands to millions level in extremely low-temperature and well-isolated systems. Here, we create multipartite entanglement of billions of motional atoms in a quantum memory at room temperature, and certify the genuine entanglement via $M$-separability witness associated with photon statistics. The information contained in a single photon is found strongly correlated with the excitation shared by the motional atoms, which intrinsically address the large system and therefore stimulate the multipartite entanglement. Remarkably, our heralded and quantum memory built-in entanglement generation allows us to directly observe the dynamic evolution of entanglement depth and further to reveal the effects of decoherence. Our results verify the existence of genuine multipartite entanglement among billions of motional atoms at ambient condition, significantly extending the boundary of the accessible scale of entanglement. Besides probing the quantum-to-classical transition in an entirely new realm, the developed abilities of manipulating such a large-scale entanglement may enhance a wide spectrum of applications for emerging quantum technologies.\\
\end{sciabstract}
\subsection*{Introduction.}
Quantum technologies, incorporating quantum entanglement~\cite{EPR_1935} into communication~\cite{np_QCm_Gisin, QIPC_2005}, simulation~\cite{S. Lloyd_Science, J.Zhang_Nature, K. Eckert_NPhysics, P. Hauke_NPhysics, X.-L. Qi_NPhysics}, computation~\cite{prl_cluster-QCp, science_OQC, nature_QCp} and metrology~\cite{review_quantum metrology}, exert great advantages beyond classical approaches. For a large-scale multipartite entangled systems, the dimension of Hilbert space will be exponentially expanded as entangled particles increase, which inspires novel approaches of quantum computing or direct simulation for classically intractable problems~\cite{QIPC_2005,nature_QCp}. The ability to access large-scale and more practical multipartite entanglement has been regarded as a benchmark for quantum information processing, like the road map towards quantum supremacy~\cite{quantum supremacy}. However, decoherence resulting from strong internal interactions and coupling with environment makes entanglement fragile, which also limits the expansion of the scale of multipartite entanglement, especially reaching the level of macroscopic physical systems. 

So far, significant experimental progresses have been made in realizing different classes of multipartite entanglement in different artificially-engineered quantum systems. Greenberger-Horne-Zeilinger state, A well-known multipartite entanglement that shares the correlation on the creation and annihilation excited on all the particles, has already been generated at a scale up to 12 qubits in superconducting systems ~\cite{PRL.122.110501}, 18 qubits in photonic systems ~\cite{PRL.120.260502}, and 20 qubits in ions systems ~\cite{PRX.8.021012}. The exponentially low efficiency in simultaneously detecting many particles restricts the achievable scale. An interesting way to enhance the collective correlation is to create twin Fock entanglement state in a Bose-Einstein condensate (BEC) system through quantum phase transitions ~\cite{science_BEC}. W state is another representative multipartite entanglement that shares the correlation of the creation excited in one particle and annihilation in all other particles. The requirement of detecting only one excitation for W state is free from the exponential inefficiency of coincidence measurement, therefore can be more easily achieved at large scale, especially in atomic ensembles~\cite{science_40atoms,nature_680atoms,nature_3000atoms,NC_million-atoms}. 

The achieved large-scale entanglement states, however, have to be prepared and detected in the systems that are maintained at extremely low temperature and well isolated with environment to eliminate decoherence effects. The decoherence and noise induced by the motion and collision of room-temperature atoms are apparently harmful~\cite{nPhys_msmemory, collision_JWPan}, and therefore were avoided in purpose in previous endeavors of observing large-scale entanglement. Though being more challenging, it would be more desirable to explore whether large-scale entanglement can exist in ambient condition and shared by many more motional atoms, not only for the fundamental interest of probing the boundary of quantum to classical transition, but also for future real-life quantum technologies.

Here, we experimentally demonstrate a multipartite entanglement of billions of motional atoms in a quantum memory operated at room temperature. The multipartite entanglement W state in a hot atomic vapor cell is heralded by registering a Stokes photon through far off-resonance spontaneous Raman scattering (SRS). In order to certify and quantify the entanglement scale, we convert the shared excitation of W state into an anti-Stokes photon by applying another interrogation pulse, and reveal a entanglement depth up to billions of atoms by the witness constructed with the correlated photon statistics. The far off-resonance configuration endorses the broadband feature of our scheme allowing to be operated at a high data rate. Furthermore, our heralded and quantum memory built-in fashion of entanglement generation allows us to directly observe the dynamic evolution of entanglement depth in a dissipative environment. 

\subsection*{Experimental implement and results.}
To herald multipartite entanglement W state, we adopt the SRS regime as proposed in the Duan-Lukin-Cirac-Zoller (DLCZ) protocol originally aiming at realizing applicable quantum repeaters ~\cite{nature_DLCZ}. However, we have to conceive a far off-resonance scheme to avoid the huge fluorescence noise in room-temperature atomic ensemble ~\cite{Dou_npj,Dou_FORD,Pang_hybrid}, which does not exist in cold ensembles and diamonds~\cite{Chou_cold,science_diamond}. The energy levels of the $\Lambda $-type configuration are shown in Figure 1a. This process will generate the product state of correlated photon-atoms pairs, which can be expressed as~\cite{science_diamond}
\begin{equation}\label{eq1}
\left | \psi _{s} \right \rangle= \left [ 1+\varepsilon_{s} S^{\dagger}a^{\dagger}\right ]\left | vac \right \rangle
\end{equation} 
where $\varepsilon_{s}$ is the excitation probability of Stokes photon, $\left |vac\right \rangle=\left |vac_{opt}\right \rangle\bigotimes \left |vac_{ato}\right \rangle$ is the initial product state of photon-atoms system, $S$ and $a$ are the annihilation operators of spin wave and Stokes photon, respectively. Here, we set the intensity of control light pulse so weak that the excitation probability $\varepsilon_{s}$ is much smaller than unity. Therefore, we can ignore the higher-order terms in the creating of  $\left | \psi _{s} \right \rangle$ with extremly small probability ~\cite{nature_DLCZ, PRL_kimble}. With the creation operators acting on the initial state of atomic ensemble, the $W$ state is written as~\cite{nPhys_msmemory,Dou_FORD} $\left| W_{1}\right\rangle {\rm{ = }}\frac{1}{{\sqrt N }}\sum\limits_{j = 1}^N {{e^{i\Delta{\vec k}\cdot {{\vec r}_j}}}} \left| {{g_1}{g_2}...{s_j}...{g_N}} \right\rangle $, where $N$ is the number of involving atoms, $\Delta{\vec k}$ is the wave-vector of spin wave, ${\vec r}_j$ is the position information of $j_{th}$ excited atom. This generating process of entanglement is shown in Figure 1b, and the W state can be heralded through the detection of one scattering Stokes photon. $\left| W_{1}\right\rangle$ contains only one excitation shared by all motional atoms illustrated in Figure 1c, where every atom posesses the equal probability being excited with spin up.

It is inevitable that the SRS in the generating process may produce high-order excitations with a comparably low probability, but such terms would change the structure of our desired multipartite entanglement. The entangled ensemble with more than one excitation event can be generally expressed as $\left| W_{2}\right\rangle {\rm{=}}\sqrt{\frac{2}{N(N-1)}}\sum\limits_{i<j}^N {{e^{i\Delta{\vec k}\cdot\left({{\vec r}_j+{\vec r}_i}\right)}}}\left|{{g_1}{g_2}...{s_i}...{s_j}...{g_N}}\right\rangle$,  and higher-order events have negligible contributions. To certify and qualify the W state, we need to apply another optical probe pulse to convert the shared single excitation in atomic entanglement state into an anti-Stokes photon, as shown in Figure 1a and Figure 1b. In order to obtain the information of entanglement depth, we analyze the photon number statistics of the correlated Stokes and anti-Stokes photons via a Hanbury Brown-Twiss interferometer (see Methods). Due to the decoherence effects, as Figure 1d shows, the atomic ensemble with high-order excitations evolves to several subgroups, where each part shares only single excitation~\cite {NC_million-atoms}. Suppose that the whole atomic ensemble is in a pure state, containing $M$ separable parts, which can be expressed in a product form $\left |\psi  \right \rangle=\left |\psi _{1}\right \rangle\otimes \left |\psi _{2}\right \rangle...\otimes \left |\psi _{M}  \right \rangle$, where $M$ indicates the number of separable subgroups, $\left |\psi _{i}\right \rangle(i=1,...,M)$ represents each separable group that may contain individual multipartite entanglement, while different subgroups are independent from the others (see Methods). Then, we can define entanglement depth as $D=N/M$, with $N$ is the number of total atoms participating in interaction~\cite {NC_million-atoms}. 

In order to quantify the multipartite entanglement, we adopt the entanglement witnesses incorporated with photon number statistics of the correlated photons pair. Such witness is efficient especially when the vacuum component of the state is dominant~\cite{NC_million-atoms}. For each given $M$ value, we ought to determine the lower bound of the entanglement state with $D$ particles entangled (see Methods). The witness operator can be expressed as
\begin{equation}\label{eq5}
\omega ^{M}=\left |W_{2} \right \rangle\left \langle W_{2} \right |-p_{2}^{bound}(p_{1},M)
\end{equation}
where the two key parameters $p_{1}$ and $p_{2}$ are projective probabilities in the forms of $\left | \left \langle W_{1}|\psi  \right \rangle \right |^{2}$ , $\left | \left \langle W_{2}|\psi  \right \rangle \right |^{2}$, and $p_{2}^{bound}(p_{1},M)$ stands for the theoretical minimal value of $p_{2}$ under the condition of fixed $p_{1}$ and $M$ value. For the density matrix $\rho$ of experimental state, $tr(\rho \omega ^{M} ) < 0$ means that the entanglement depth is at least $D=N/M$. In actual experiments, $p_{1}$ should be defined as the conditional probability of detecting a correlated anti-Stokes photon with a heralded forward Stokes photon, and the probability $p_{2}$ stands for the probability of two excitation events, which is deduced by the autocorrelation function $g_{AS_{1}-AS_{2}|S}^{(2)}={2p_{2}}/{p_{1}^{2}}$ measured by a Hanbury Brown-Twiss interferometer. 

The decoherence effects can be revealed by the observation of entanglement depth's evolution via adjusting the delay of optical probe pulse in our quantum memory built-in configuration. The $p_{1}$ can be influenced by the retrieval efficiency of quantum memory, the photon loss of the channels and detectors. Therefore, the final experimental datas should be handled as the following two stages: the raw measured datas and the processed datas after subtracting the loss of the channels and detectors. The latter reflects the actual entanglement state at the moment just after applying the probe pulse. The $M$ values of entanglement state evolving with storage time are showed in Figure 2a, where the energy of the light pulses is 225 $pJ$. The data point below the boundary curve of fixed $M$ value indicates that there is an entanglement depth at least $N/M$. Our experimental datas show that the number of entanglement subgroups increases as the memory time elapsing, i.e. the decoherence effects caused by the thermal motions of atoms~\cite{nPhys_msmemory,Dou_FORD} will tremendously influence the structure of entanglement, which is consistent with the physical picture depicted in Figure 1d. 

During the process of verifying the existence of entanglement, the collective enhancement effect contributes to the transducing of Stokes photon due to the phase coherence of $W$ state \cite{nature_DLCZ}. The variation of $p_{1}$ with the delay of the probe pulse is shown in Figure 3a. Due to the decoherence of phase mainly resulting from the motions of warm atoms, the effects of collective enhancement become deteriorative, which results in the exponential decay of $p_{1}$. What is more, we measure the cross-correlation between the correlated photons and the autocorrelation of the retrieved anti-Stokes photon as shown in Figure 3b. The degrade of quantum correlation and single-photon characteristic implies the variation of the structure of multipartite entanglement according to the relation of $p_{2}$ and $g_{AS_{1}-AS_{2}|S}^{(2)}$ (see Methods), which are consistent with the deduced $M$ values in Figure 2. Our results well exhibit the transition from quantum to classical in multipartite entanglement of billions of motional atoms heralded by single photon.

From another perspective, we can demonstrate how the entanglement depth varies with the delay time. In order to determine the informations of entanglement depth, the number of total caesium atoms involved in the interaction is the key parameter that should be measured precisely. The atomic density and total atomic number can be obtained by fitting the measured transmission rates of light with different frequency according to the absorption model (see Methods). The results show that there are nearly at least 8.85 billion motional atoms sharing the one excitation constituting the W state. As is shown in Figure 3c, despite the fast decrease of entanglement depth resulting from the increased noise and the destructive effects of decoherence, there are still considerable entanglement depth in the warm atomic ensemble after storing for the time of microseconds level. 

It is also accessible to manipulate the size of large-scale entanglement state in the macroscopic ensemble by changing our experimental parameters. There are several elements that will influence the entanglement depth, such as the beam waist, the detuning, and the energy of addressing light. For the far off-resonance DLCZ protocol, we have chosen a ``sweet point" for the detuning, which has been experimentally demonstrated to have the lowest unconditional noise~\cite{Dou_npj,Dou_FORD}. As for the beam waist, it is not appropriate to utilize too large beam waist to provide enough addressing energy. Note that there is no problem for using larger beam waist given that strong addressing light is equipped, whose advantage is that there will be more atoms involved in the creation of entanglement state. In our experiment, we choose a beam waist of $100 \mu m$ for providing a sufficient excitation rate. What dominantly influences the $p_1$ probability in our experiment is the energy of the addressing light pulse, which determines the excitation probability of the Stokes photon during the SRS process. 

The $M$ values of multipartite entanglement created by different pulse energies, 115.5$pJ$, 225$pJ$, 330$pJ$ respectively, are shown in Figure 4a. The relation among entanglement depth and excitation probability, light pulse energy is also analyzed and shown in Figure 4b. The results show that the stronger light pulse energy has a higher $p_{1}$ probability because of the higher converting efficiency of $W$ state, but has a smaller $M$ value, which indicates that the structure of multipartite entanglement has not been deteriorated by noises. Actually, for conveniently and efficiently evaluating the scale of entanglement, the aforementioned definition of entanglement depth only delivers a lower bound of the actual entanglement scale~\cite{NC_million-atoms}, since the scale of genuine entanglement depth should be associated with the largest size of all subgroups. Thus, it is reasonable that the same lower bound of entanglement depth is observed with different excitation probabilities.  

\subsection*{Discussion and Conclusion.}
In summary, we have experimentally demonstrated a multipartite entanglement of billions of motional atoms heralded by single photon. With the quantum memory built-in and broadband capacities, we have efficiently displayed the dynamical evolution of entanglement depth and decoherence effects, as well as showing the transition from quantum to classical in the realm of multipartite entanglement. Our work has certified that quantum entanglement can be observed in a macroscopic room-temperature atomic ensemble with motional atoms and demonstrated the accessibility of feasible manipulations of entanglement depth, which greatly expanded the bound of operating large-scale multipartite entanglement and may have potential applications for future quantum information science and technologies. 

Creating a larger-scale multipartite entanglement beyond billions of atoms is possible, a larger beam waist and stronger energy of light pulse will be helpful with the prerequisite of well controlled levels of noise. What's more, the larger beam waist can mitigate the detrimental effect of decoherence brought by the thermal motions of atoms, since there is a broader space to prevent warm atoms from escaping from the interaction area, which leads to a longer lifetime of multipartite entanglement. Recent works also show that the anti-relaxation coating of vapor cell will preserve the coherence for longer time~\cite{balaba_coating,polsik_DLCZ}, which may be beneficial for improving the maintenance of the heralded multipartite entanglement. Remarkably, recent proposals and experimental developments about transferring the single collective excitation of electronic spins to noble-gas nuclear spins by spin exchange regime may exceedingly prolong the lifetimes of the W state even up to several hours~\cite{prl_nuclear spin,ofer_noble-gas}.

The $W$ state with phase informations encoded in billions of atoms exists in the form of a spin wave, which resemble a tremendous networked quantum sensors with entanglement between each elements. The phase informations of spin wave are not only related to the position informations of motional atoms, but also sensitive to some other physical parameters related to atomic internal states, like magnetic field~\cite{interfermeter_prl}, which makes the $W$ state become a promising candidate for quantum sensing. Due to the collective enhancement effects in the readout of spin wave, the huge scale will become an advantage in enhanced metrology. The nonclassical correlations contained in the $W$ state among huge entangled particles may endow the meteorological gain over classical states~\cite{PRL_Wgain}, such as being beneficial to interferometry measurement for beating standard quantum limit~\cite{review_quantum metrology}. Interestingly, $W$ state is also robust for the purpose of metrology, because the remaining particles are still entangled while one particle is trace out.  Furthermore, the multipartite entanglement constructed between these quantum sensors may significantly enhance the precision of multiparameter estimation~\cite{PRL_MPE}. 

\subsection*{Acknowledgments.}
The authors thank Jian-Wei Pan for helpful discussions. This research was supported by the National Key R\&D Program of China (2019YFA0308700, 2017YFA0303700), the National Natural Science Foundation of China (61734005, 11761141014, 11690033), the Science and Technology Commission of Shanghai Municipality (STCSM) (17JC1400403), and the Shanghai Municipal Education Commission (SMEC) (2017-01-07-00-02- E00049). X.-M.J. acknowledges additional support from a Shanghai talent program.\\

\subsection*{Methods}
\paragraph*{Experimental details:} The $^{133}$Cs cell is 75-mm-long and is placed into a magnetic field shielding, which has been filled 10 $Torr$ Ne buffer gas to alleviate collisions between cesium atoms. In order to get a large optical depth, which means there are more atoms participating in the creation of entanglement, the $^{133}$Cs cell is heated to 61$^{\circ}$C. To generate high-speed light pulse with enough intensity shown in Figure 1, we have developed a system to satisfy the needs of tunable central frequency, broad bandwidth, and more importantly, control pulse generated in a programable fashion. We have also established a frequency locked system to stabilize the W state during the creating and certifying process. It should be noticed that our collinear scheme makes the correlated Stokes and anti-Stokes photons being coaxial propagating under the phase-matching condition, therefore six home-built Fabry-P\'erot cavities with high performance are employed to split and retrieve Stokes and anti-Stokes photons. As for single cavity, the transmission rate of each cavity reaches around $95\%$ , and the extinction rate of each cavity for signal and noise is up to $500:1$. 

\paragraph*{Witness for $M$-separability:} In the certifying process, we apply an optical pulse to interrogate the state of atomic ensemble by analyzing the correlated photons statistics. Firstly, we assume that the atomic state can be described in a pure state, which can be decomposed into a $M$-separable form like equation $\left |\psi  \right \rangle$ in the main text. Due to the imperfect experimental conditions and decoherence, the representation of state in every independent subgroup may be a superposition of many possible states\cite{Duan_25memory(2017)}. Since we consider at most two excitations in the SRS process, the specific form of each state in the subgroup can be expressed in following state
\begin{equation}\label{eq1}
\left |\phi _{i} \right \rangle=a_{i}\left |W _{0} \right \rangle+b_{i}\left |W _{1} \right \rangle+c_{i}\left |W _{2} \right \rangle
\end{equation}
where $\left |W_{1} \right \rangle$, $\left |W_{2} \right \rangle$ are the Dicke states in each subgroup; $\left |W_{0} \right \rangle$ is a vacuum state. Thus, the whole state of ensemble is the product of all subgroups:
\begin{equation}\label{eq2}
\left |\Psi  \right \rangle=\bigotimes_{i=1}^{M}a_{i}\left |W _{0} \right \rangle+b_{i}\left |W _{1} \right \rangle+c_{i}\left |W _{2} \right \rangle
\end{equation}
The two probability $p_{1}$, $p_{2}$ can be calculated specifically,
\begin{equation}\label{eq3}
p_{1}=\frac{\left |\prod_{i=1}^{M}a_{i}  \right |^{2}}{M}\left | \sum_{i} \frac{b_{i}}{a_{i}}\right |^{2}
\end{equation}
\begin{equation}\label{eq4}
p_{2}=\frac{\left |\prod_{i=1}^{M}a_{i}  \right |^{2}}{M^{2}(1-\frac{1}{N})}\left | \sqrt{2}\sum_{i<j} \frac{b_{i}b_{j}}{a_{i}a_{j}}+\sqrt{1-\frac{1}{D}}\sum_{i} \frac{c_{i}}{a_{i}}\right |^{2}
\end{equation}
where $D=\frac{N}{M}$ is the entanglement depth. Actually, the entanglement depth should be defined as the largest scale of all the entangled subgroups, however, we can take ${N}/{M}$ as the representation of entanglement depth to avoiding get vary large subgroup size\cite{NC_million-atoms}.   

In order to determine the entanglement depth in experiment, we need to determine the lower bound for $p_{2}$ with a fixed $M$ value. Obviously, we need the probability $p_{2}$ as low as possible in actual experiment, which means that the fidelity of target $W_{1}$ state is high. The bound can be calculated by
\begin{equation}\label{eq5}
p_{2}^{bound}(p_{1},M)=min{\left \{ p_{2}|\psi :p_{1}, M = const \right \}}
\end{equation}
Note that there are constraints between the coefficients of superposition in equation (3), $\left | a_{i} \right |^{2}+\left | b_{i} \right |^{2}+\left | c_{i} \right |^{2} \leq 1 $, and we can take the approximation $|a_{i}|^{2}+|b_{i}|^{2}+|c_{i}|^{2}= 1$ owing to the neglectable higher excitations. Utilizing the Lagrange multiplier method deduced in the supplementary notes of \cite{NC_million-atoms}, the conclusion is that the symmetric solution gives the global minimal value of $ p_{2}$ for $M\leq5$. This symmetric solution requires that $a_{i}=a, b_{i}=b, c=-\sqrt{1-a^{2}-b^{2}}$. In this condition, the optimal values for $p_{1}, p_{2}$ are the solution as follows
\begin{equation}\label{eq6}
p_{1}^{sym}=Ma^{2M-2}b^{2}
\end{equation}
\begin{equation}\label{eq7}
p_{2}^{sym}=a^{2M}(\frac{1}{\sqrt{2}}(M-1)\frac{b^{2}}{a^{2}}+\frac{c}{a})^{2}
\end{equation} 
The final form of function $p_{2}^{bound}(p_{1},M)$ is,
\begin{equation}\label{eq8}
p_{2}^{bound}(p_{1},M)=a^{2M}[\frac{1}{\sqrt{2}}\frac{p_{1}(M-1)}{M}a^{-2M}-\frac{1}{a}\sqrt{1-a^{2}-\frac{p_{1}}{M}a^{2-2M}}]^{2}
\end{equation}
For fixed $p_{1}$ and $ M$, we need to calculate minimal value of $p_{2}$ with $a\in (0,1)$. The theoretical bound of equation (10) can be obtained by take all values of $p_{1}\in (0,1)$, which is shown in Figure 2a and Figure 3b with different $M$ values.

\paragraph*{Number of atoms involved in the creation of multipartite entanglement:} Theoretically, the transmission rate of probe light passing through the atomic ensemble has the following form~\cite{phdOxford_Sprague,absorption_model}
\begin{equation}\label{eq6}
T(\omega )=exp\left \{-\frac{2\pi nkLd^{2}}{h\varepsilon _{0}}\sum_{i=1}^{3}S_{i}l_{i}(\omega ) \right \}
\end{equation}
where  $n$ is the density of atoms, $k$ is the wave vector of probe light, $L$ is the length of our vapor cell, $d$ is the reduced dipole matrix element, $S_{i}$ is the strength of relative coupling from the hyperfine level $F=3$ of the
ground state to $F'=2,3,4$ in the excited state\cite{Cs_Data}, $l_{i}(\omega )$ is the normalization lineshape. For more precisely fitting, the normalization lineshape $l_{i}(\omega )$ should be considered as Vigot lineshape~\cite{absorption_model}. More details about the theoretical absorption model and experimental fitting are in the supplementary notes. According to the fitting coefficients, the number density of atoms in the $^{133}$Cs cell  is nearly $1.21\times 10^{18} m^{-3}$. 

To determine how many atoms participating in the interaction, we need to know the volume of interaction area illuminated by light in the creating and certifying process, which can be evaluated with the volume of Gaussian light within the cell. The light beam used to generate multipartite entanglement is Gaussian beam, whose amplitude of electric field has the spatial dependence in following form
\begin{equation}\label{eq6}
|E(r)|=\frac{E_{0}}{\sqrt{1+\frac{z^{2}}{z_{w}^{2}}}}e^{-\frac{x^{2}+y^{2}}{W_{w}^{2}(1+\frac{z^{2}}{z_{w}^{2}})}}
\end{equation}
where $z_{w}$ is Rayleigh length of laser beam, $W_{w}$ is the beam waist. Here, we only define the effective area of interaction by the amplitude decrease to ${1}/{10}$ of central magnitude in the Gaussian beam. Actually, the witness applied to certify our $W$ state doesn't require equal amplitude for the entanglement state in $\left| W_{1}\right\rangle$ ~\cite{NC_million-atoms}, therefore more atoms can be taken into calculations owing to the extension of Gaussian beam's intensity. Therefore, the illuminated volume can be calculated as,
\begin{equation}\label{eq6}
V=\pi ln10\int_{-\frac{l}{2}}^{\frac{l}{2}}W_{w}^{2}(1+\frac{z^{2}}{z_{w}^{2}})dz
\end{equation}
where $l=75.3\times10^{-3}m$ is the length of cesium cell, and beam waist $W_{w}=100\times10^{-6}m$, Rayleigh length $z_{w}=3.69\times10^{-2}m$. Finally, the total number of atoms $N$ involved into the creation of entanglement is $N=nV=8.85\times 10^{9}$.

\clearpage

\clearpage

\begin{figure}[htbp]
	\centering
	\includegraphics[width=1\linewidth]{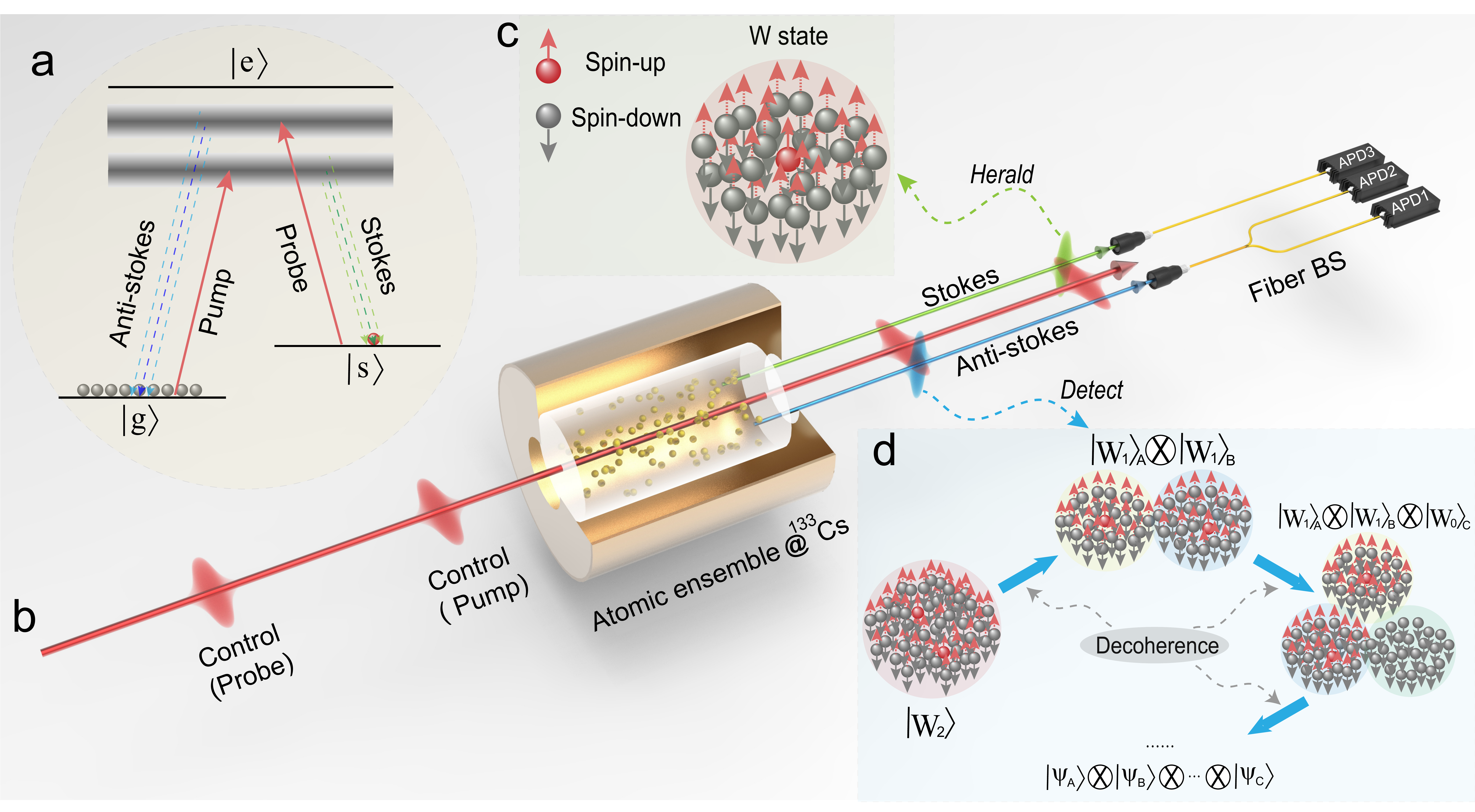}
	\caption{\textbf{The schematic diagram of creation and verification of multipartite entanglement.}  \textbf{a.} The energy levels of creating and verifying W state. Solid lines represent three-level $\Lambda$-type configuration of atoms, two ground states label $\left|g\right\rangle$ ($6S_{1/2}, F=3$) with electronic spin down and $\left|s\right\rangle$ ($6S_{1/2}, F=4$) with electronic spin up, which are hyperfine ground states of cesium atoms (splitting is $\Delta g = 9.2{\rm{GHz}}$) ; excited state labels $\left| {\rm{e}} \right\rangle $ ($6P_{3/2}, F^{'}=2,3,4,5$). The shaded area between energy levels represent broad virtual energy levels induced by the short pump and probe laser pulse (2ns). \textbf{b.} The experimental scheme of creating and certifying the multipartite entanglement. The Hanbury Brown-Twiss interferometer is used for analyzing the statistics of the correlated Stokes and anti-Stokes photons, which can further reveal the informations of entanglement depth. \textbf{c.} The representation of the dashed arrow means that every atom has equal probability to be spin up or down, which is the main feature of the W state. \textbf{d.} The decoherence effects may change the structure of the multipartite entanglement. We take the two excitations event as example to show the evolutions of entanglement depth. The distributed cesium atoms with different colors illustrate the entanglement distribution with several subgroups. The set of same colored atoms is genuine multipartite entanglement, while two sets with different colors do not have the relationship of entanglement.}
	\label{f1}
\end{figure}

\clearpage

\begin{figure}[htbp]
	\centering
	\includegraphics[width=1.0\linewidth]{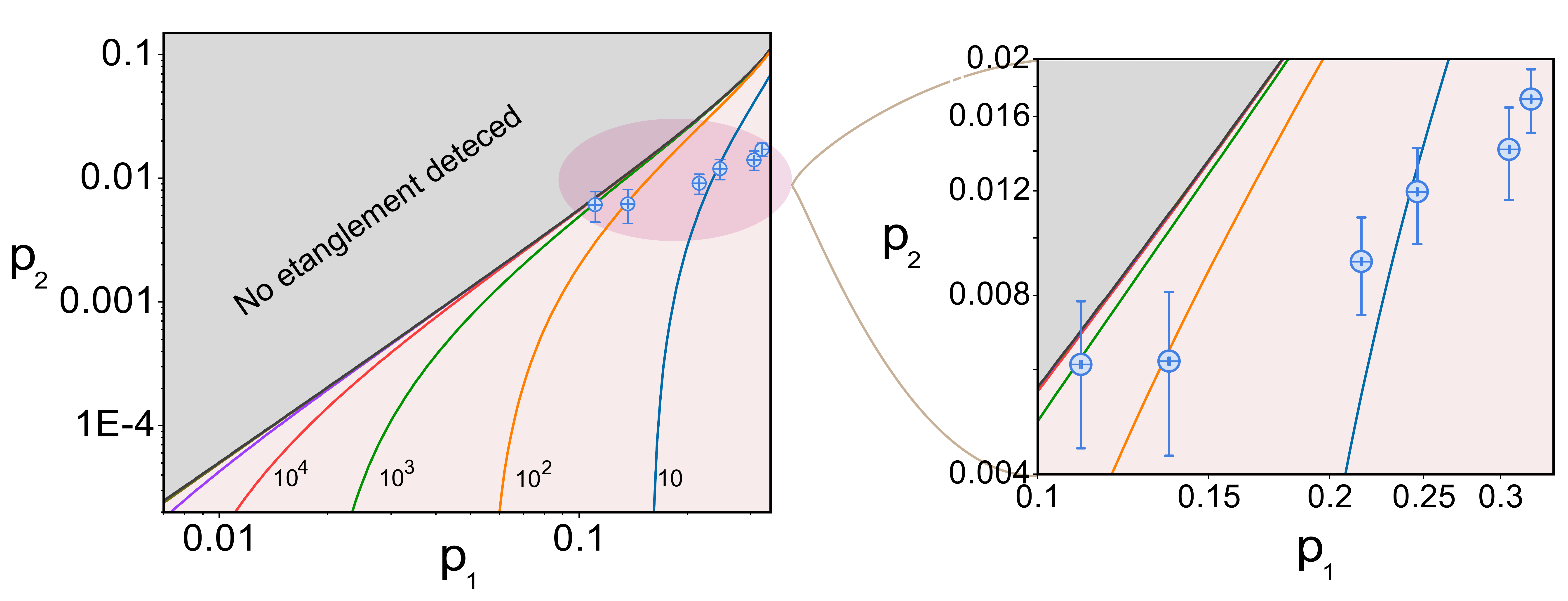}
	\caption{\textbf{Dynamical evolution of the multipartite entanglement.} The verification results of witness  and dynamical evolution of $M$ values with different storage time. The different $M$ values of corresponding theoretical lower bound curves of witness are 10 to $10^{6}$ from right to left. The $M$ values for experimental datas are 5, 6, 10, 14, 92 and 1000 in the right subgragh. Error bars are derived by Poisson distribution of photon number statistics from avalanche photo diodes.}
	\label{f2}
\end{figure}

\clearpage

\begin{figure}[htbp]
	\centering
	\includegraphics[width=1.0\linewidth]{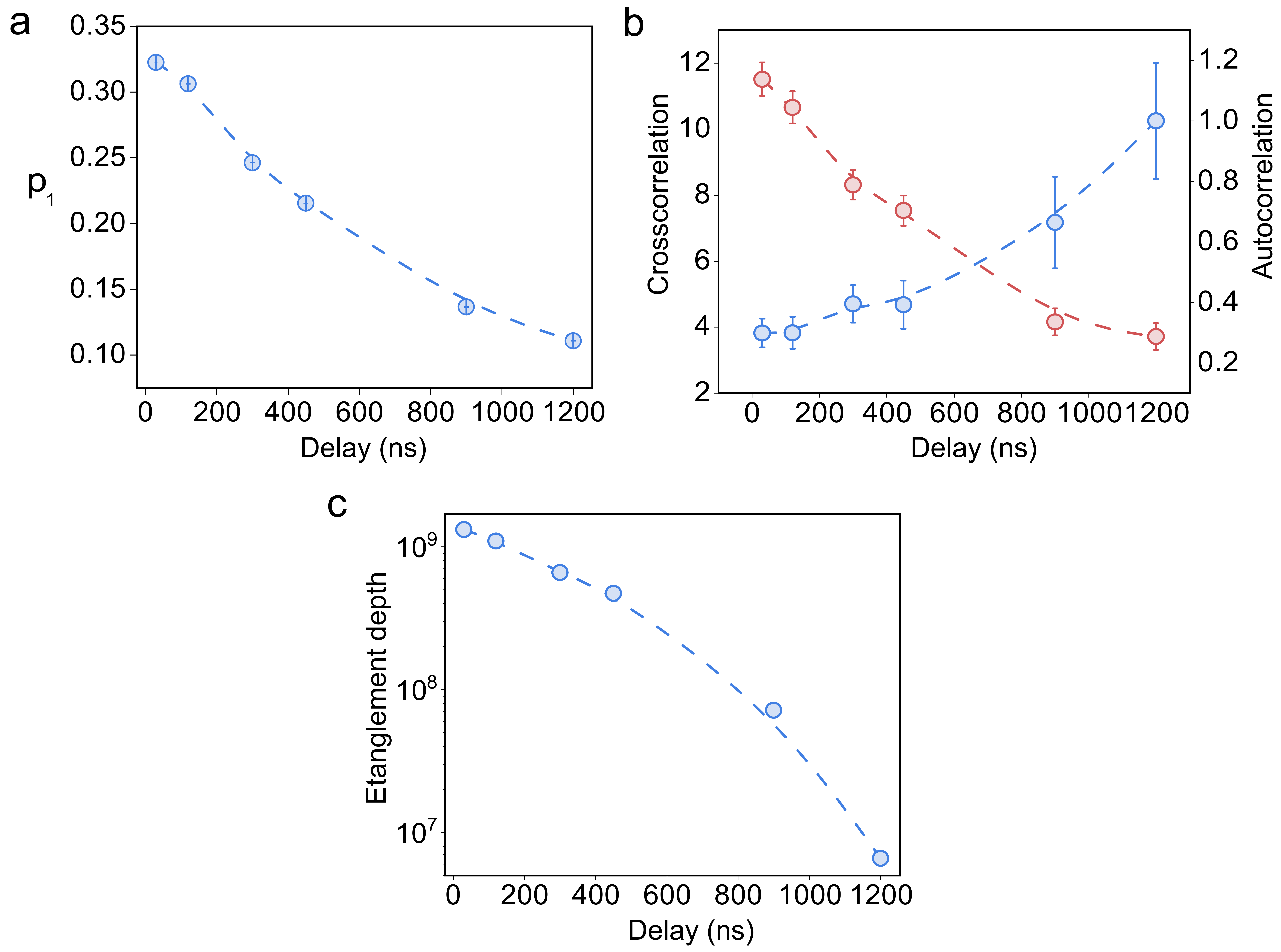}
	\caption{\textbf{ Decoherence of the multipartite entanglement.} \textbf{a.} The variations of  $p_{1}$ probabilities. \textbf{b.} The crosscorrelation $g_{S-AS}^{(2)}$ between correlated stokes photon and anti-stokes photon is in the left side (as pink dots show); the autocorrelation $g_{AS_{1}-AS_{2}|S}^{(2)}$ of the retrieved anti-Stokes photon by a heralded Stokes photon is in the right side (as blue dots show). Error bars are derived by Poisson distribution of photon number statistics from avalanche photo diodes. \textbf{c.} The evolution of entanglement depth varies with the storage time of quantum memory.}
		\label{f3}
\end{figure}

\clearpage

\begin{figure}[htbp]
	\centering
	\includegraphics[width=0.85\linewidth]{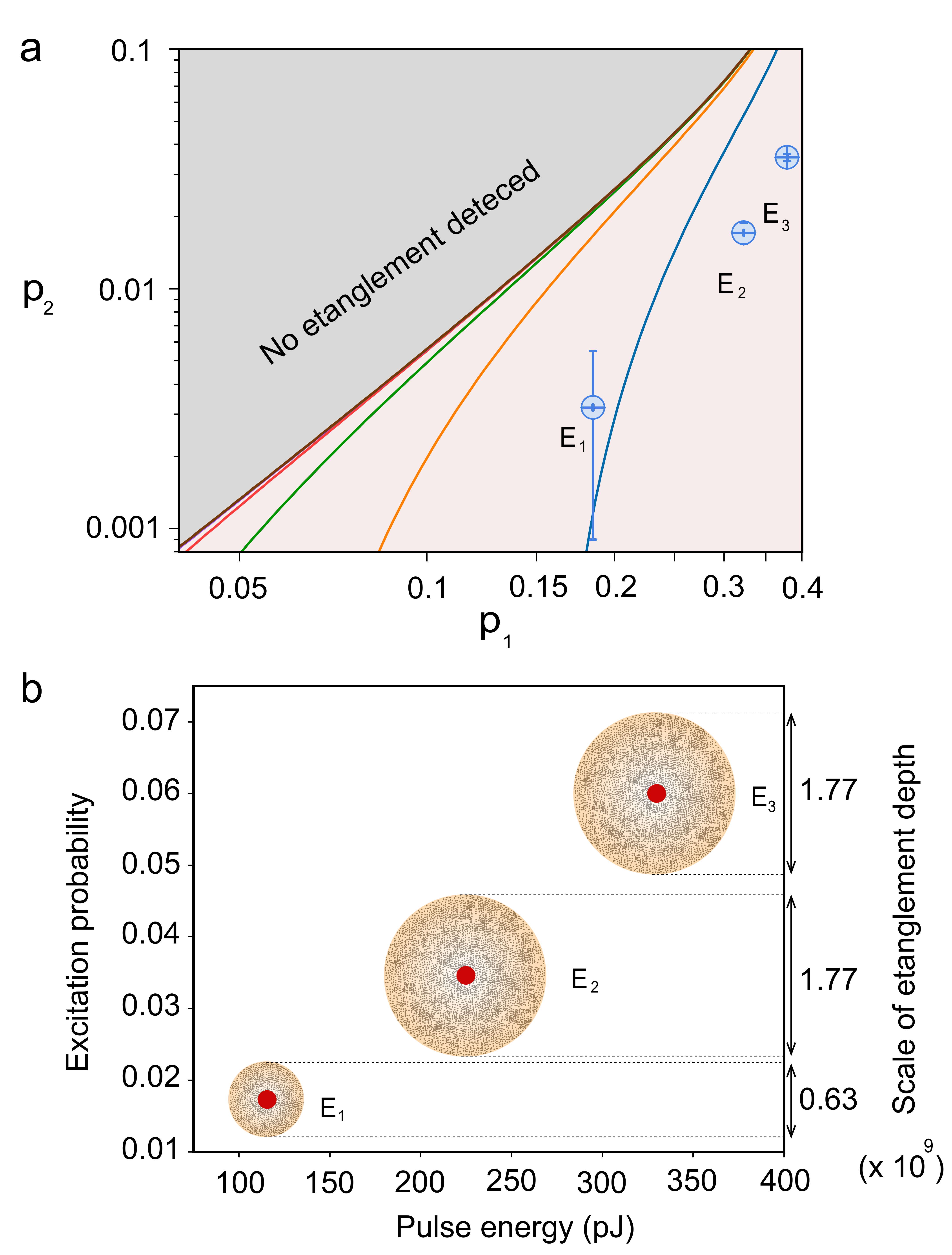}
	\caption{\textbf{ Measured entanglement depth with different excitation probabilities.} \textbf{a.} The $M$ values with different excitation probabilities are 14, 5, 5, and the corresponding energy are $E_{1}=115.5$ $pJ$, $E_{2}=225$ $pJ$, $E_{3}=330$ $pJ$. The time interval between creation and verification light pulse is 30 ns. Error bars are derived by Poisson distribution of photon number statistics from avalanche photo diodes. \textbf{b.} The visulized entanglement depth for comparason with different excitation probabilities. The red dots represent the relations of the excitation probability and pulse energy, and the size of circle taking the red dot as center stands for the scale of entanglement depth.}
		\label{f3}
\end{figure}

\clearpage
\bigskip
\section*{\large Supplementary Materials: Multipartite Entanglement of Billions of Motional Atoms Heralded by Single Photon}
\setcounter{figure}{0}
\setcounter{table}{0}
\setcounter{equation}{0}
\renewcommand{\figurename}{Supplementary Figure}
\renewcommand{\tablename}{Supplementary Table}

\renewcommand{\thetable}{\arabic{table}}
\renewcommand{\theequation}{{S}\arabic{equation}}

\section*{\large I. Number density of atoms in the atomic ensemble }  

For determine number density of our vapor cell, we have measured the absorption curve of sweeping frequency. Therefore, it is necessary to build up a theoretical model to fit the absorption curve. The absorption of light beam through a atomic vapor is described by the Beer-Lambert law: 
\begin{equation}\label{eq1}
I(z)=I_{0}exp[{-\alpha (\nu,T_{0} )z}]
\end{equation}
where $I_{0}$ is the beam intensity in the entrance of cell, and absorption coefficient $\alpha (\nu,T_{0} )$is dependent on frequency $\nu$ and temperature  $T_{0}$. During the experiment, we use beam pulse, whose intensity is weak enough, so the absorption coefficient is independent of intensity. However, there is not only single transitions existing, but many absorption transitions occurring in Cs atoms. As a result, the transmission $T(\omega )$ should be extended as: 
\begin{equation}\label{eq2}
T(\omega)=exp[-(\sum\alpha _{i})L ] 
\end{equation}   
where $\sum S_{i}\alpha _{i}$ is total absorption coefficient, $L$ is length of vapor cell. In our system, considering the selection rules of transitions,  the possible transition mainly occurs between hyperfine level $F=3$ of the ground state$(6S_{1/2})$ to $F'=2,3,4$ in the excited state$(6P_{3/2})$. So, these absorption coefficients can be denoted as $\alpha _{32}$, $\alpha _{33}$ and $\alpha _{34}$, finally the actual transmission can be expressed in following form:
\begin{equation}\label{eq3}
T(\omega)=exp[-(\alpha _{32}+\alpha _{33}+\alpha _{34})L ] 
\end{equation} 

The relative strengths of transition $S_{32}$, $S _{33}$ and $S _{34}$ has been regarded as usually used constants of database, which can be obtained in \cite{Cs_Data}. The absorption coefficients $\alpha _{32}$, $\alpha _{33}$ and $\alpha _{34}$, are related to the electric susceptibilities $\chi _{32}$, $\chi _{33}$ and $\chi _{34}$ \cite{absorption_model}:  
\begin{equation}\label{eq4}
\chi _{i}(\omega ) =S_{i}d^{2}N\frac{2\pi }{h\varepsilon _{0}}l_{i}(\omega ) 
\end{equation}  
where $i$ stands for notations of $32$, $33$and $34$; $S_{i}$ is the relative strength of different transitions, which can also be denoted as $S_{32}$, $S _{33}$ and $S _{34}$; d is the matrix element of transition; N is the number density of atoms; $l_{i }(\omega )$ is the lineshape factor of transition $i$, which can be derived from the optical Bloch equations of two-level atom\cite{Loudon}. And then, we can get the expression of absorption coefficient:
\begin{equation}\label{eq5}
\alpha _{i}(\omega )=k\chi _{i}(\Delta )=kS_{i}d^{2}N\frac{2\pi }{h\varepsilon _{0}}l_{i }(\omega ) 
\end{equation} 
where k is the wave number of probe light, since the range of sweeping frequency in measurement is small, the wave number can be viewed as constant.

Taking equation $(5)$ into $(3)$, we can get the specific form of transimission: 
\begin{equation}  
T(\omega )=exp\left \{-\frac{2\pi NkLd^{2}}{h\varepsilon _{0}}(S_{32}l_{32 }(\omega ) +S_{33}l_{33 }(\omega )+S_{34}l_{34 }(\omega )) \right \}
\end{equation} 
Note that the lineshape factor of transition $l_{i }(\omega )$ should take Doppler lineshape into consideration for better agreeing well with the experimental datas.Therefore, considering Doppler effects and collision broadening mechanism, the total lineshape is Vigot type\cite{Loudon, absorption_model}, which is the convolution of Lorentz lineshape and Doppler lineshape in the form of equation $(7)$.
\begin{equation}\label{eq7}
l_{i}(\omega )=[f_{\Gamma }\otimes g_{\sigma }](\omega )=\int_{-\infty }^{+\infty }f_{\Gamma }(\Delta -\delta _{\omega })g_{\sigma }(\delta _{\omega })d\delta _{\omega }
\end{equation}
where $\delta _{\omega }=kv_{z}=\frac{\omega}{c}v_{z}$ is Doppler shift of frequency, $v_{z}$ is the velocity of atom along the direction of wave vector $k$, $c$ is the velocuty of light in vaccum; $\Delta=\omega-\omega_{i}$ is the detune of probe light from the resonant frequency; $f_{\Gamma }(\omega)$, $g_{\sigma }(\omega)$ are normalized Lorentz lineshape and Doppler lineshape,  respectively, whose expressions as equation$(8), (9)$.
\begin{equation}\label{eq8}
f_{\Gamma }(\omega )=\frac{\frac{\Gamma}{2\pi } }{\Delta ^{2}+(\frac{\Gamma }{2})^{2}}\end{equation} 
\begin{equation}\label{eq9}
g_{\sigma }(\omega )=\frac{1}{\sigma \sqrt{2\pi }}exp[-\frac{1}{2}(\frac{\delta_{\omega} }{\sigma  })^{2}]\end{equation}
where $\Gamma$ is the linewidth of Lorentz lineshape, and $\sigma$ is the linewidth of Doppler lineshape. Inserting equation$(8), (9)$ into $(7)$, we can get the function of Voigt lineshape:
\begin{equation}\label{eq10}
l_{i}(\omega )=\frac{2\omega }{\pi \sqrt{2\pi }\Gamma \sigma c}\int_{-\infty }^{+\infty }\frac{1}{1+(\frac{\omega -\omega _{i}-\frac{\omega }{c}v_{z}}{\frac{\Gamma }{2}})^{2}}exp[-\frac{1}{2}(\frac{v_{z}}{\frac{c}{\omega }\sigma })^{2}]dv_{z}
\end{equation}
To simplify the form of equation $(10)$, we can make a variable replacement $\omega ^{'}=\omega -\frac{\omega }{c}v_{z}-\omega_{i}$, and get the expression of Voigt lineshape again:
\begin{equation}\label{eq11}
l_{i}(\omega )=\frac{2}{\pi \sqrt{2\pi }\Gamma \sigma }\int_{-\infty }^{+\infty }\frac{1}{1+(\frac{\omega ^{'}}{ \frac{\Gamma}{2}})^{2}}exp[-\frac{1}{2}(\frac{\Delta- \omega^{'}}{\sigma })^{2}]d\omega ^{'}
\end{equation} 
where $\Delta=  \omega-\omega_{i}$.

To obtain the specific expression of total transmission, we can replace $l_{i}(\omega)$ in equation $(6)$ with equation $(11)$ :
\begin{equation}\label{eq12}
\begin{split} 
T(\omega )=&exp\left \{ \frac{4d^{2}NkL}{\sqrt{2\pi }h\varepsilon _{0}\Gamma \sigma }[ \int_{-\infty }^{+\infty }(S_{32}\cdot exp[-\frac{1}{2}(\frac{\omega-\omega _{32} -\omega ^{'}}{\sigma })^{2}] \right. \\ 
&\left. + S_{33}\cdot exp[-\frac{1}{2}(\frac{\omega-\omega _{33} -\omega ^{'}}{\sigma })^{2}] +S_{34}\cdot exp[-\frac{1}{2}(\frac{\omega-\omega _{34} -\omega ^{'}}{\sigma })^{2}])\times \frac{1}{1+(\frac{\omega ^{'}}{\frac{\Gamma }{2}})^{2}} d\omega ^{'}] \right \}
\end{split}
\end{equation}
And we can simplify the fitting equatuion$(12)$ by define fitting parameters $k_{1}$, $k_{2}$ and $k_{3}$, so we can get fitting equation $(13)$.
\begin{equation}\label{eq13}
\begin{split}
T(\omega )=&exp\left \{ k_{1}[ \int_{-\infty }^{+\infty }(S_{32}\cdot exp[-\frac{1}{2}(\frac{\omega-\omega _{32} -\omega ^{'}}{k_{2} })^{2}]+\right. \\ 
&\left.S_{33}\cdot exp[-\frac{1}{2}(\frac{\omega-\omega _{33} -\omega ^{'}}{k_{2} })^{2}]  +S_{34}\cdot exp[-\frac{1}{2}(\frac{\omega-\omega _{34} -\omega ^{'}}{k_{2} })^{2}])\times \frac{1}{1+(\frac{\omega ^{'}}{k_{3}})^{2}} d\omega ^{'}] \right \}
\end{split}
\end{equation}
where 
\begin{equation}\label{eq14}
k_{1}=\frac{4d^{2}NkL}{\sqrt{2\pi }h\varepsilon _{0}\Gamma \sigma }
\end{equation}
\begin{equation}\label{eq15}
k_{2}= \sigma
\end{equation}
\begin{equation}\label{eq16}
k_{3}=\frac{\Gamma }{2}
\end{equation}

\begin{figure}
	\centering
	\includegraphics[width=0.95\columnwidth]{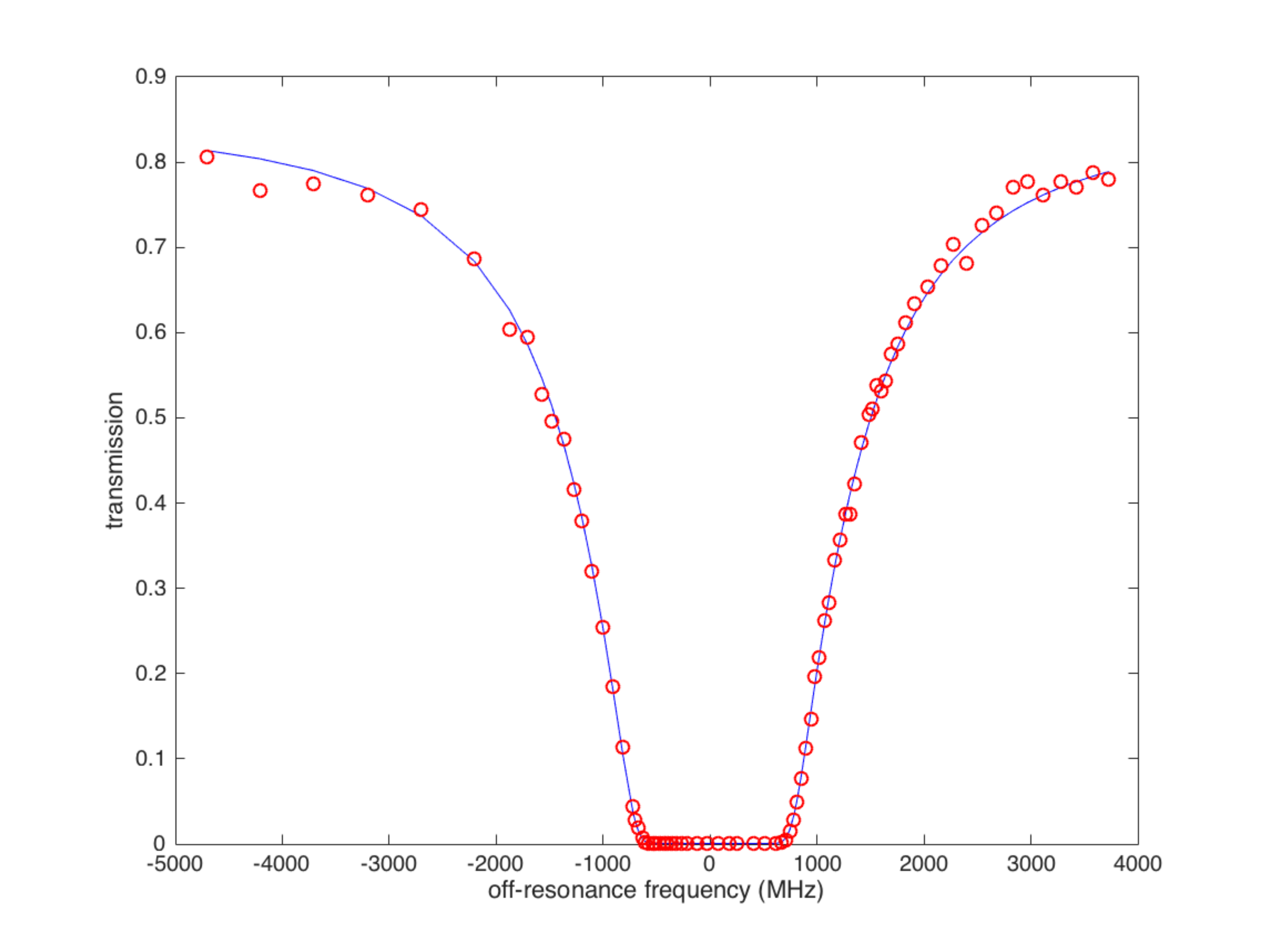}\\
	\caption{\textbf{Fitting of transmission rate curve}. The fitting result of transmission rate of the weak probe light with different detuning.}
	\label{f2}
\end{figure}

Due to the actual experimental conditions, like inevitable noise and the errors of measurement,  the measured transmission maybe a little different from theoretical values. We can add two extra parameters to transform fitting equation $(13)$ into $T(\omega )^{'}$.
\begin{equation}\label{eq17}
T(\omega )^{'}=k_{4}T(\omega )+k_{5}
\end{equation}
The finally fitting result is showed in Fig s1. The key fitting parameters $k_{1}$, $k_{2}$ and $k_{3}$ are $0.6039$, $189.9895$ and $61.0204$. The unit of parameters $k_{2}$ and $k_{3}$ is $MHz$ in actual experiment. From equation $(14)-(16)$, we can get the relation between number density $N$ and fitting parameters:
\begin{equation}\label{eq17}
N=\frac{\sqrt{2\pi }h\varepsilon _{0}k_{1}k_{2}k_{3}}{2d^{2}kL}
\end{equation}
where matrix element $d$ take the value of effective far-detuned dipole moment $2.1923\times 10^{-29} C\cdot m$\cite{Cs_Data}, $k=\frac{\omega_{0}}{c}$ and $\omega_{0}=2\pi \cdot 351.726 THz$ is the frequency of cesium $D_{2}$ transition, the length of our cesium cell is $75.3mm$. Finally, the number density of atoms is evaluated as $1.2133\times 10^{18} m^{-3}$.

\clearpage

\end{document}